\documentclass[aps,prl,reprint,superscriptaddress]{revtex4-1}

\pdfoutput=1
\usepackage[margin=1in]{geometry}
\usepackage{amsmath}
\usepackage[caption=false]{subfig}
\usepackage[hidelinks]{hyperref}
\usepackage{graphicx}
\begin{document}
\raggedbottom
%\preprint{}

%------------------------------
% Title and Author Information
%------------------------------

\title{Examining the Transition from Multiphoton to Optical-Field Photoemission From Silicon Nanostructures}


\author{Phillip D. Keathley}
%\email[]{pdkeat2@mit.edu}
%\thanks{}
\affiliation{Research Laboratory of Electronics, Massachusetts Institute of Technology, 77 Massachusetts Ave., Cambridge, MA 02139}
\author{William P. Putnam}
\affiliation{Research Laboratory of Electronics, Massachusetts Institute of Technology, 77 Massachusetts Ave., Cambridge, MA 02139}
\affiliation{Department of Physics and The Hamburg Center for Ultrafast Imaging, University of Hamburg, Luruper Chau{\ss}ee 149, 22761 Hamburg, Germany}
\affiliation{NG Next, Northrop Grumman Corporation, 1 Space Park Blvd., Redondo Beach, CA 90278, USA}
\author{Guillaume Laurent}
\affiliation{Research Laboratory of Electronics, Massachusetts Institute of Technology, 77 Massachusetts Ave., Cambridge, MA 02139}
\affiliation{Physics Department, Auburn University, Auburn, AL 36849}
\author{Luis F. Vel\'asquez-Garc\'ia}
\affiliation{Microsystems Technology Laboratories, Massachusetts Institute of Technology, 77 Massachusetts Ave., Cambridge, MA 02139}
\author{Franz X. K\"artner}
\affiliation{Research Laboratory of Electronics, Massachusetts Institute of Technology, 77 Massachusetts Ave., Cambridge, MA 02139}
\affiliation{Department of Physics and The Hamburg Center for Ultrafast Imaging, University of Hamburg, Luruper Chau{\ss}ee 149, 22761 Hamburg, Germany}
\affiliation{Center for Free-Electron Laser Science, Deutsches Elektronen-Synchrotron (DESY), Notkestra{\ss}e 85, 22607 Hamburg, Germany}


\date{\today}

\begin{abstract}
We perform a detailed experimental and theoretical study of the transition from multiphoton to optical-field photoemission from n-doped, single-crystal silicon nanotips.  Around this transition, we measure an enhanced emission rate as well as intensity-dependent structure in the photoelectron yield from the illuminated nanostructures.  Numerically solving the time-dependent Schr\"odinger equation (TDSE), we demonstrate that the excess emission derives from the build-up of standing electronic wavepackets near the surface of the silicon, and the intensity dependent structure in this transition results from the increased ponderomotive potential and channel closing effects.  By way of time-dependent perturbation theory (TDPT), we then show that the visibility of intensity dependent structure, the transition rate from multiphoton to optical-field emission, and scaling rate at high intensities are all consistent with a narrow band of ground-state energies near the conduction band dominating the emission process in silicon.  These results highlight the importance of considering both coherent electron wavepacket dynamics at the emitter surface as well as the ground-state energy distribution when interpreting strong-field photoemission from solids.   
\end{abstract}

%\keywords{}

\maketitle

%-------------
% Letter Body
The transition from multiphoton (MP) to optical-field (OF) emission (sometimes referred to as optical tunneling emission) is of key importance to high harmonic generation in solids and gases, attosecond science, and lightwave electronic devices~\cite{sivis_tailored_2017,ghimire_observation_2011,goulielmakis_attosecond_2007,schiffrin_optical-field-induced_2013,eckle_attosecond_2008,yudin_nonadiabatic_2001,zimmermann_unified_2017}.  The Keldysh parameter, $\gamma$, defined as the ratio of the tunneling time to the duration of the incident laser cycle~\cite{keldysh_ionization_1965}, is often used to delineate these two regions, where $\gamma \ll 1$ indicates OF emission, and $\gamma \gg 1$ indicates MP emission.  While these regimes have separate, intuitive pictures, the transition region holds properties of both MP and OF emission~\cite{yudin_nonadiabatic_2001}, and is rich with physics such as ponderomotive shifting of the bands, light-induced states, and channel-switching~\cite{zimmermann_unified_2017,bardsley_ac_1989,kruit_ac_1983}.

Channel closing and resonant behavior in the MP-OF transition have been observed in atoms~\cite{zimmermann_unified_2017,paulus_channel-closing-induced_2001}, however for solid cathodes a smooth, featureless rollover in yield is typically observed~\cite{bormann_tip-enhanced_2010,anisimov_nonlinear_1977,farkas_influence_1972,putnam_optical-field-controlled_2016, hobbs_high-density_2014, hobbs_high-yield_2014}. Given that solids have lower work functions than gases, it should be easier to resolve features in the yield scaling due to the reduced scaling rate.  However, other complications arise due to optical damage, heating~\cite{kealhofer_ultrafast_2012,anisimov_nonlinear_1977,riffe_femtosecond_1993}, space charge effects, and more complicated band structures and ground state distributions.

To avoid optical damage, space charge, and thermal effects, subwavelength structures can be used to achieve highly localized electric field enhancements (greater than a factor of ten in $\text{nm}^3$ volumes), over sufficiently short pulse durations (few to tens of fs)~\cite{bormann_tip-enhanced_2010,swanwick_nanostructured_2014}.  This enables optical field emission deep into the tunneling regime from solids using relatively low average power~\cite{bormann_tip-enhanced_2010,racz_measurement_2017}. Applications include near-field ultrafast electron microscopy~\cite{cocker_tracking_2016}, lightwave driven electronic devices~\cite{rybka_sub-cycle_2016, putnam_optical-field-controlled_2016,schiffrin_optical-field-induced_2013,goulielmakis_attosecond_2007}, as well as spatially and temporally structured photocathodes and photoinjectors~\cite{keathley_strong-field_2012,swanwick_nanostructured_2014,racz_measurement_2017,dombi_ultrafast_2013,li_surface-plasmon_2013}.  

Here we present a detailed experimental and theoretical study of the intensity-dependent photoelectron yield from silicon nanotips in the MP-OF transition region.  The experimental results exhibit an enhanced emission rate and intensity dependent features in the yield scaling.  Using analysis based on integration of the time-dependent Schr\"{o}dinger equation (TDSE) and time-dependent perturbation theory (TDPT), we show that the enhanced emission is due to the formation of standing electronic wavepackets near the emitter surface and their interaction with the oscillating potential. Furthermore, the structure is due the ponderomotive shifting of the continuum threshold leading to decoupling of the standing wavepackets and, eventually, channel closure.  Finally, an analysis of the effects of the ground state energy distribution shows that the visibility of intensity dependent structure in the yield scaling, the MP-OF transition rate, and the final slope of the the emission scaling rate are all consistent with a narrow band of ground-state energies near the conduction band dominating emission. 

\begin{figure}
\includegraphics[width=0.45\textwidth]{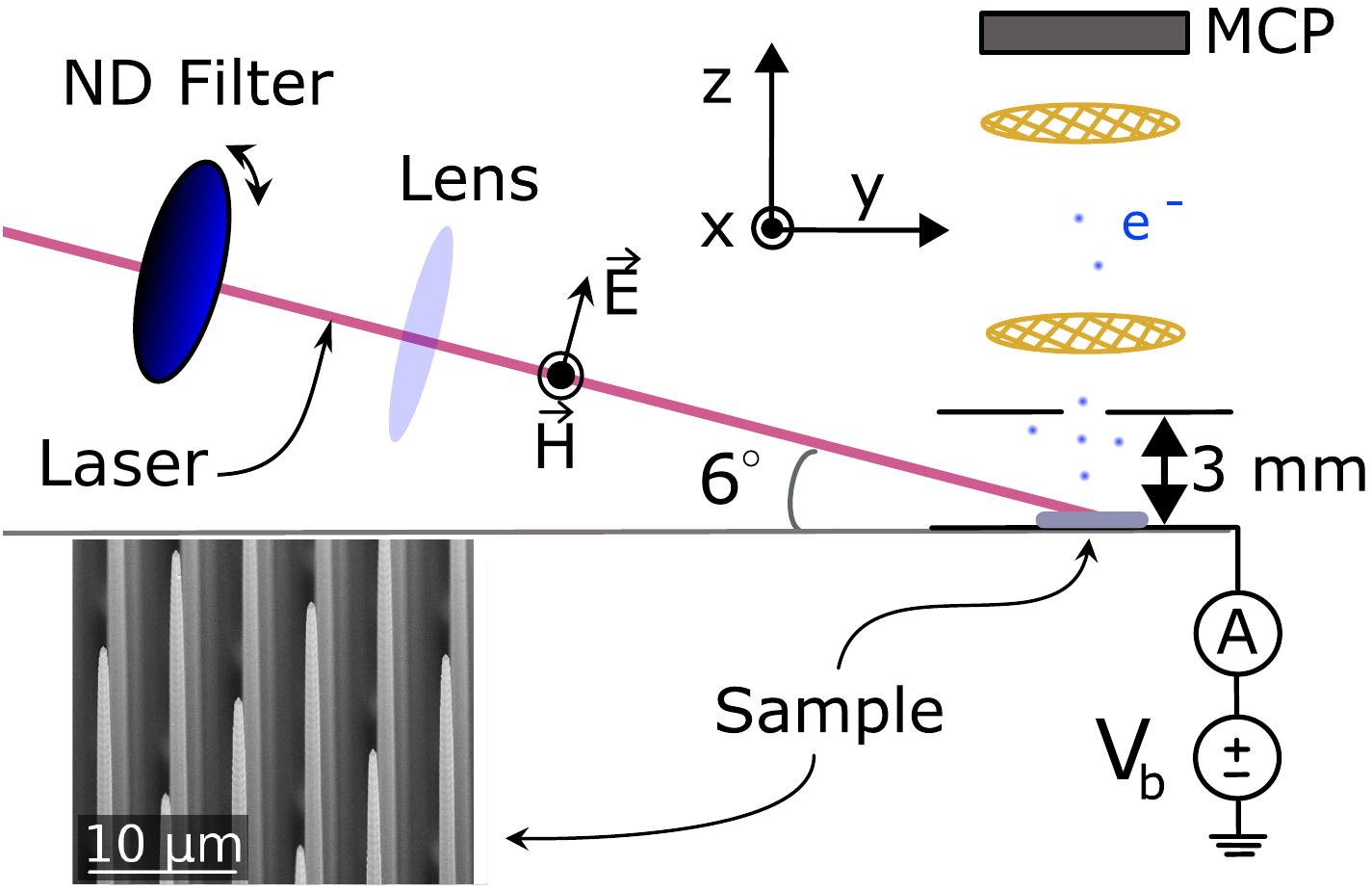}%
\caption{Experimental setup.  The top shows a schematic of the optical configuration with the sample in place.  The sample and holder are biased to $V_b = -10\text{\ V}$ relative to the chamber, and the current leaving the sample is measured through a transimpedance amplifier. The lower-left is a scanning electron micrograph of the tip array. \label{fig:system_setup}}
\end{figure}

The experimental setup is shown in Fig.~\ref{fig:system_setup}.  Pulses having a repetition rate of 3 kHz, average pulse duration of $\approx 55 \text{\ fs}$ full width at half maximum (FWHM), and center wavelength of 800 nm were focused onto an array of tips at a grazing incident angle of $\approx 6^{\circ}$. The pulse duration was measured at the tip surface by performing interferometric autocorrelations using the emitted current, and, using a CCD camera, the spot size at the focus was measured to be roughly 200 $\mu$m FWHM.  The incident pulse energies ranged up to $2.5\text{\ }\mu\text{J}$ and were adjusted with a variable neutral density filter.  The measured spatial profile, incident pulse energies and temporal duration were used to calculate peak field intensities before enhancement. 

%%%%%%%%%%%%%%%%%%%%%%%%%%%%%%%%%%%%%%%%%%%%%%%
% System Setup
%
% System configuration and tip image.
%%%%%%%%%%%%%%%%%%%%%%%%%%%%%%%%%%%%%%%%%%%%%%%

The tips were etched from a $\langle100\rangle$ n-doped, single-crystal silicon wafer, with a doping concentrationn on the order of $10^{15}$ cm${}^{-3}$, putting the Fermi level $\approx 4.3$ eV below the vacuum level. The spacing between the tips was 10 $\mu$m, and each tip had an average radius of curvature of $\approx 10 \text{\ nm}$ at its apex.  The photoemission current was measured through the substrate, to which a negative bias of 10 V was applied relative to the anode.  The anode also served as the entrance aperture of a time of flight electron spectrometer (TOF). All measurements were performed in high vacuum ($10^{-8}\text{\ Torr}$).  

%%%%%%%%%%%%%%%%%%%%%%%%%%%%%%%%%%%%%%%%%%%%%%%
% Current Scaling Figure
%
% Shows experimental and theoretical current 
% scaling as a function of intensity.
%%%%%%%%%%%%%%%%%%%%%%%%%%%%%%%%%%%%%%%%%%%%%%%
\begin{figure}
\includegraphics[width=\columnwidth]{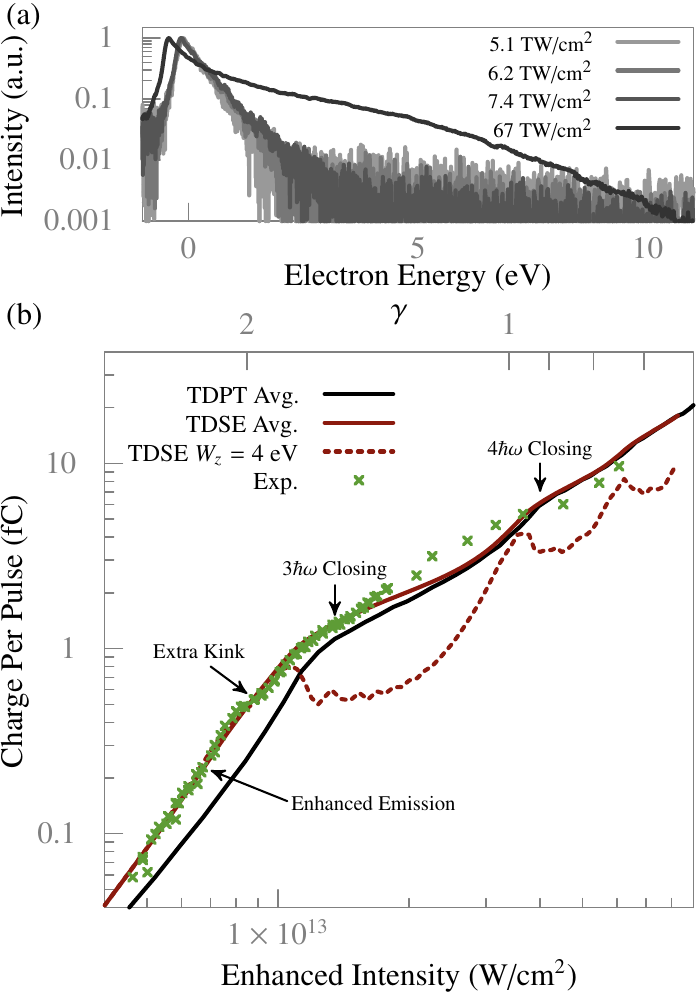}%
\caption{Experimental results. (a) Electron energy spectra measured at select intensities. (b) Current scaling as a function of peak intensity. The experimental datapoints (green crosses) are from three scans: two fine scans near the kink (one increasing in intensity, the other decreasing) and a coarse scan going to higher pulse energies.  Theoretical scaling curves include the ground state and spatially averaged TDPT (solid black) and TDSE results (solid red), as well as the TDSE result for a discrete ground state energy level $W_z = 4\text{\ eV}$ (dashed red). \label{fig:current_scaling}}
\end{figure}

The experimental results are shown in Fig.~\ref{fig:current_scaling}. In Fig.~\ref{fig:current_scaling}a we show electron energy spectra measured at select peak intensities at the emitter surface. At low intensities, the shape of the normalized electron energy spectra show almost no change as a function of intensity, despite $\approx 5 \times$ increase in yield. However, at high optical intensities, the electron energy spectra develop a broad plateau that extends beyond $10\text{\ eV}$; this plateau is consistent with the OF emission regime in which the emitted electrons can achieve large kinetic energies by rescattering off the emitter surface~\cite{dombi_ultrafast_2013,herink_field-driven_2012}.  In Fig.~\ref{fig:current_scaling}b, we show the photoelectron yield as a function of peak intensity at the emitter surface.  The sudden change, or kink, in emission near $1\times10^{13}\text{\ W/cm}{}^2$ is consistent with a transition from MP to OF emission from the tips~\cite{bormann_tip-enhanced_2010,farkas_influence_1972,irvine_ponderomotive_2005}. (Note that in prior work we have found that DC bias dependence and pump-probe measurements rule out effects due to space charge and thermionic emission~\cite{swanwick_nanostructured_2014}).   

The intensity dependent structure in the yield scaling during the MP-OF transition was repeatedly observed at the same incident intensities over days of measurement. The data shown in Fig.~\ref{fig:current_scaling}b were collected over three consecutive sweeps: first a coarse sweep of increasing pulse intensities, then a fine sweep of increasing and then decreasing intensity values.  The narrow distribution of the datapoints indicates a high consistency in laser pointing, pulse energy, and pulse duration across all three scans. We should also note that the enhanced intensity values come from fitting the curve to the TDSE model results (solid, red line) via a least squares method.  The field enhancement factor providing the best fit was found to be $\approx{24}$.  

In the following, we briefly overview how total charge yield was modeled using the TDSE and TDPT.  The calculation results are then discussed and compared to the experimental data.  For both methods, an initial state was taken to be an effectively free electron with energy $W_z = U - k_z^2/2$ below the continuum (i.e. the vacuum level), where $U$ is the step-potential height, and $k_z$ is the momentum in $z$ toward the barrier (atomic units are used throughout unless stated otherwise). For conduction band electrons in Si, we take $U = \chi = 4.05$ eV (where $\chi$ is the electron affinity). 

The TDSE and TDPT were then used to calculate the emission probability which is defined as the ratio of the outgoing charge density emitted into the continuum, i.e. the adjacent vacuum, to the incoming current density toward the step-boundary.  The emission probability is expressed as $\Gamma(W_z, \mathcal{E}_0)$, $\mathcal{E}_0$ being the peak electric field at the tip surface.  By normalizing the initial state wavefunctions such that $|\psi| = 1$ at the step-boundary, we can write $\Gamma(W_z, \mathcal{E}_0) = g(W_z)C(W_z, \mathcal{E}_0)$, where $g(W_z) = 2 \sqrt{2(U - W_z)}/U$ accounts for the current density toward the step, and $C(W_z, \mathcal{E}_0)$ for the outgoing charge density (see Supplementary Material for more details). 

For the TDSE, a Crank-Nicolson scheme with discrete transparent boundary conditions was used~\cite{yalunin_strong-field_2011}, and for the TDPT we account for direct ionization from a step-boundary using the strong-field approximation~\cite{swanwick_nanostructured_2014,yalunin_strong-field_2011,milosevic_above-threshold_2006}  (further details provided in the Supplementary Material). After obtaining $\Gamma(W_z, \mathcal{E}_0)$, we calculated the total emitted charge by integrating over a ground state energy distribution (each ground state energy level was assumed to contribute to the total current incoherently). Similar integration has been described by numerous authors~\cite{fowler_electron_1928,yalunin_strong-field_2011}.  The total electron yield can be written as
\begin{equation}
Q(\mathcal{E}_0) \propto \int_{-\infty}^{U} \mathrm{d}W_z g(W_z) F(W_z) C(W_z,\mathcal{E}_0) \mbox{.}
\end{equation}
The term $F(W_z) = \ln \big [ 1 + \exp \lbrace \beta (W_f + W_z - U) \rbrace \big ]/(2\pi^2\beta)$ comes from the projection of the electron momentum distribution onto the z axis, where $\beta = 1/k_BT$ is the inverse thermal energy, and $W_f$ is the Fermi level relative to the bottom of the step.  For all calculations, the temperature was taken to be $300\text{\ K}$.

Following the calculation procedure described above, we spatially averaged over the beam spot (accounting for the variable intensity throughout the focused spot) and arrived at the solid black and red curve in Fig.~\ref{fig:current_scaling}b for the TDPT and TDSE respectively. Only the intensity enhancement and an overall multiplicative constant were fit. The emission curve predicted by the TDSE is in remarkable agreement with experimental data, especially in the transition region from MP to OF emission. A single emission scaling curve from $W_z = 4.0$ eV (dashed red) is shown for comparison (without any spatial averaging) and provides an excellent fit at low intensities through the first kink of the experimental data but has a much steeper drop in emission after the channel closing than what is observed experimentally.  This perhaps indicates that the method of averaging, incorporation of the decaying field, or exact ground state distribution does not perfectly match the experiment and needs further investigation.  Going to higher intensities, the dips in the experimental yield coincide with both the $3\hbar\omega$ and $4\hbar\omega$ channel closing intensities predicted by both the TDPT and TDSE.  

%%%%%%%%%%%%%%%%%%%%%%%%%%%%%%%%%%%%%%%%%%%%%%%
% TDSE Wave function Dynamics
%%%%%%%%%%%%%%%%%%%%%%%%%%%%%%%%%%%%%%%%%%%%%%%
\begin{figure*} \includegraphics[width=\textwidth]{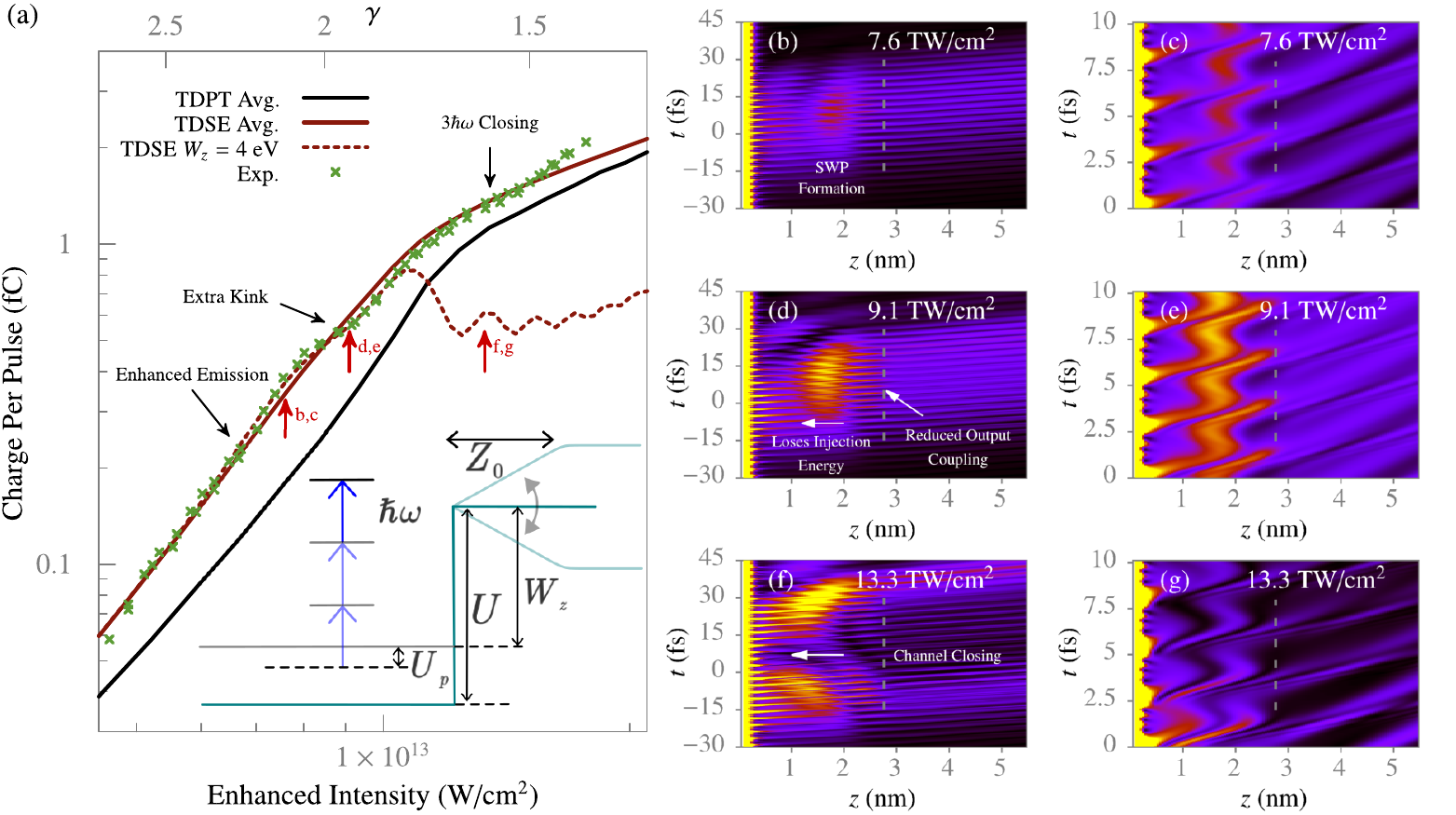}%
\caption{TDSE results. (a) Fine scan results near the kink (experimental) plotted together with the TDSE and TDPT results (same as Fig.~\ref{fig:current_scaling}a).  The inset shows a diagram of the modeled step-potential with key parameters labeled. The red arrows indicate intensity values used for plotting the squared magnitude of the wavefunction shown in figures (b-g). In (b-g), the driving pulse peak is centered at $t = 0$ fs. The dashed line indicates the field cutoff point at $z_0$. Each row shows the same result on two timescales; for (b,c) the multiphoton region is shown; (d,e) shows the wavefunction near the first kink; and (f,g) shows the wavefunction just after the first channel closing. \label{fig:tdse_results}}
\end{figure*}

The TDSE results are examined more closely in Fig.~\ref{fig:tdse_results}.  Two key differences between the TDSE and TDPT are: (i) the field is ``turned off'' at a finite distance away from the tip surface, $z_0$, to simulate the field decay away from a sharp tip, and (ii) the emitted electrons can interact with the step-potential after emission.  These differences have the greatest effect near the MP to OF transition.  

Due to the electron reflection from the surface, forward and backward moving components of the wavefunction create standing wave packets (SWPs) in the region $0 < z < z_0$ (Fig.~\ref{fig:tdse_results}b,c). These packets are not stationary, but are moved in time by the oscillating field (Fig.~\ref{fig:tdse_results}c).  Due to the finite nature of the field decay, the SWP furthest from the emitter surface can be pushed through to the region outside the oscillating field (see Fig.~\ref{fig:tdse_results}c), which effectively dictates the ``output coupling'' of the SWP into the field free region where it can escape the surface. This leads to excess emission when compared to the TDPT, where both interaction with the surface and field decay effects have been ignored.  Additionally, note that the SWP comes to a peak at $t>0\text{\ fs}$ while the center of the driving field is at $t = 0\text{\ fs}$.  This is because the SWP is formed by the mixing of the electron wavefunction in the oscillating cavity with newly injected wavefunction amplitude; this explains why these SWPs and the resulting enhancement in emission is not observed for short driving pulses. 

At higher intensitites, i.e. at higher field strengths, the electron injection momentum is reduced due to the ponderomotive AC Stark shift of the step-barrier. With a driving frequency of $\omega$, the step-barrier shifts by $U_p = {\mathcal{E}_0}^2/(4\omega^2)$. This AC Stark shift and the resulting loss in injection momentum causes the position of the SWPs to move toward the surface as the intensity increases (Fig.~\ref{fig:tdse_results}d,e); such movement results in a reduced output coupling of the SWP and a commensurate increase in SWP intensity (Fig.~\ref{fig:tdse_results}d,e) as well as a slight kink in emission yield (see the extra kink in Fig.~\ref{fig:tdse_results}a).  

As the intensity is further increased, the electron injection momentum continues to drop due to the AC Stark shift.  The emission rate increases again before dropping suddenly due to closure of the three-photon channel when $3\hbar\omega < W_z + U_p$.  This channel closure is visualized in Fig.~\ref{fig:tdse_results}f as the SWP appears to have been sucked into the step-boundary.  This effect is also predicted by the TDPT, indicating that it is fundamental to the electron injection rather than to rescattering. As the emission is driven further into the OF regime, the TDPT and TDSE converge, showing subsequent channel closures as $U_p$ continues to rise (Fig.~\ref{fig:current_scaling}b).  Thus at higher intensities, emission is dominated by the tunneling injection rate rather than the coherent electron dynamics at the surface.  

%Think about wording here
%This perhaps indicates that the initial peak observed experimentally is due to another physical mechanism.  It is known, for instance, that resonances can also be observed as peaks in emission scaling due to the AC stark shift~\cite{tang_many_1990}.

\begin{figure}
\includegraphics{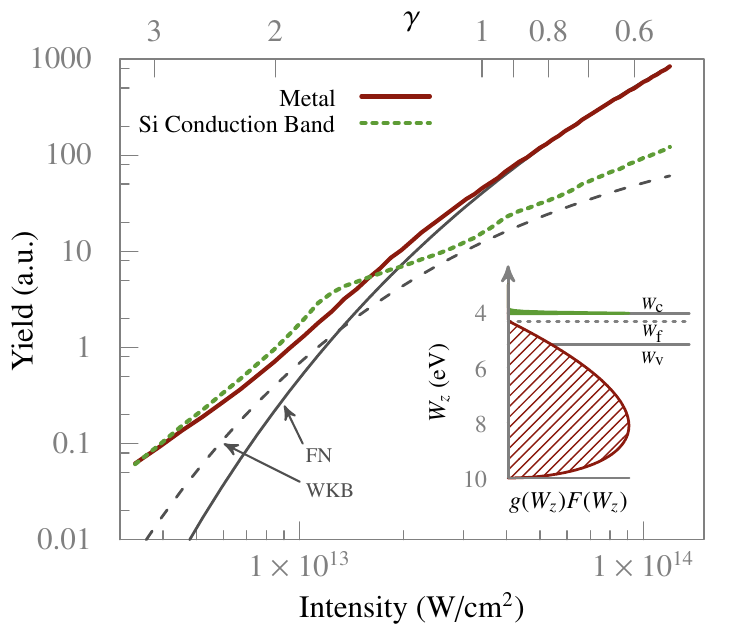}%
\caption{Comparison of emission from conduction band of Si (green) to an ideal metal (red).  In the inset, $g(W_z)F(W_z)$ for both cases are also plotted. The variables $W_c$ and $W_v$ denote the bottom of the conduction band and top of the valence band respectively.  Curves relating to the static FN (solid gray) and WKB (dashed gray) tunneling rates are shown for reference.\label{fig:metal_si_comparison}}
\end{figure}

As the TDPT provides a good fit during the initial channel closing and into the OF emission regime, and as this model is less computationally expensive, we use it to study how different ground state distributions would alter the emission curve.  First, we compare the modeled scaling curve for conduction band electrons in Si with that of an ideal metal having a step-potential height of 10 eV and the same Fermi level~\footnote{For the ideal metal, 10 eV was chosen as the step-potential height as the conduction band of metals typically lies on the order of 10 eV below the vacuum level}.  Both scaling curves along with normalized plots of $g(W_z)F(W_z)$, which relates to the electron density at the surface, are shown in Fig.~\ref{fig:metal_si_comparison}.  Note that these calculations also include spatial averaging over the beam spot.

For the metal, the emission scaling exhibits a slightly reduced slope in the multiphoton region as well as a much slower rollover into the OF emission regime.  Additionally, the emission from the metal shows a higher scaling rate at high incident intensities, and no indication of channel closures. Also, as expected, in the tunneling regime, it converges with the Fowler-Nordheim (FN) rate equation (solid gray)~\cite{fowler_electron_1928,yalunin_strong-field_2011}.  On the other hand, the scaling profile from the silicon more closely follows that of a single energy level near $W_z = 4\text{\ eV}$ due to the narrow pool of states in the conduction band.  For the silicon, the transition from MP-OF emission is sharper, with distinct $3\hbar\omega$ and $4\hbar\omega$ channel closings.  Additionally, as expected, the emission rate from the silicon at high intensities more closely follows that given by the Wentzel-Kramers-Brillouin (WKB) tunneling probability through a static triangular barrier (dashed gray)~\cite{fowler_electron_1928,yalunin_strong-field_2011}.  

\begin{figure}
\includegraphics{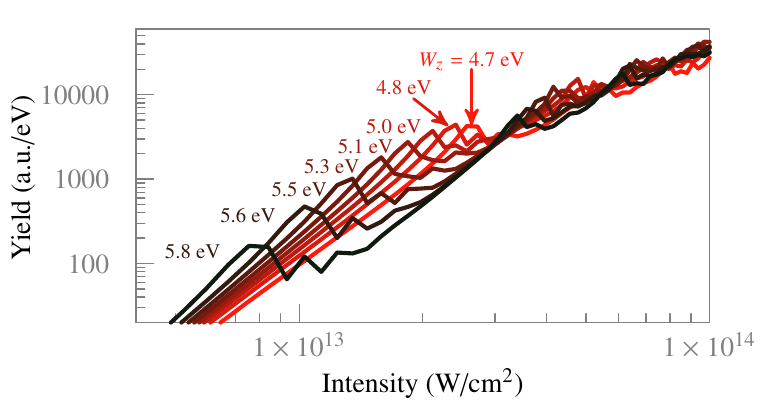}%
\caption{Electron yield contributions for select $W_z$ values as a function of intensity. \label{fig:resonance_blurring}}
\end{figure}

Lastly, to better understand how each energy level contributes to the scaling curve, the intensity scaling contributions for the ideal metal ($F(W_z)\Gamma(W_z, \mathcal{E}_0)$) are plotted for select ground state energies as a function of intensity in Fig.~\ref{fig:resonance_blurring}.  For each curve shown, there is a sharp drop in yield for the the 4-, 5-, and 6-photon channel closings.  For increasing $W_z$, the kink location moves to lower and lower intensity values.  This continues, and as $W_z$ continues to increase, eventually the first open channel becomes the 5-photon channel, then the 6-photon channel, and so forth.  Thus, for the metal, the MP-OF transition intensity is not well defined, and the channel closures are filled-in as a broader distribution of ground state levels contribute to the total emitted charge (see Supplementary Material for further discussion). 

%%%%%%%%%%%%%%%%%%%%%%%%%%%%%%%%%%%%%%%%%%%%%%%
% Metal/Si Conduction Band Comparison
%
% Shows comparison of yield scaling when using
% an idealized metal ground state distribution
% vs. Si conduction band.
%%%%%%%%%%%%%%%%%%%%%%%%%%%%%%%%%%%%%%%%%%%%%%%

%%%%%%%%%%%%%%%%%%%%%%%%%%%%%%%%%%%%%%%%%%%%%%%
% Resonance Blurring Figure
%
% Shows how resonance blurring occurs due to 
% a broad ground state energy distribution.
%%%%%%%%%%%%%%%%%%%%%%%%%%%%%%%%%%%%%%%%%%%%%%%

%Remove this...not fully supported, and beyond scope of this letter!
%To investigate further, the yield from several combinations of top-hat distributions of $g(W_z)F(W_z)$ having different amplitude were compared against the experimental data.  In the end, we found the best fit was always for a narrow band of states (~0.1-0.3 eV wide) near the conduction band edge, with some possible contribution from a narrow band of surface states lying near the conduction band.  This is consistent with the width of the slight bump in emission just before the tunneling kink being just $0.16$ eV wide as a function of $U_p$.

%Debating about putting this in...I think there are multiple factors making this very difficult.  They often work below the keldysh parameter of 1 in their work, meaning that the emission process is not probing too deep into the ground state...this coupled with a higher work function could also act to localize the contributing ground states, and lead to more structure in the outgoing electron spectrum.  I don't think I want to open this can of worms...
% It is important to note that coherent energy peaks have been observed in the longitudinal energy structure of electrons emitted from crystalline tungsten\addref{HOMMELHOFF REFS}.  While this also implies a narrow band of ground states dominating the emission process, the short pulse durations used mean that the photon energy is not well defined, which also acts to smear out any resonant features in the total emission yield\addref{Yalunin}.

In conclusion, we have experimentally observed enhanced emission and intensity dependent structure in the photoelectron yield from silicon nanotips.  Using models based on the TDSE and TDPT, we determined that the enhanced emission comes from the buildup of standing electronic wavepackets near the surface of the silicon emitters, and the structure from ponderomotive shifting of the continuum and channel closing effects. We also examined the effects of differing ground state populations on the emission scaling, and showed how these populations affect the transition rate from MP to OF emission, the visibility of intensity dependent structure, and the final scaling rate deep into the OF regime. 

Unlike atomic systems, solids provide an opportunity to engineer the underlying band structure of the emitting material, and to correspondingly tune the properties of the emitted electrons. A narrow ground state energy distribution such as that found in Si combined with the AC Stark shift provides a probe to better our understanding of strong-field emission from solids and the properties of the emitted wavepackets.  We hope future investigations of the emitted electrons' transverse energy structure will provide more information about the effects of band structure and coherent electron dynamics on strong-field emission from solids. 

%------------------
% Acknowledgements
% ------------------
\section*{Acknowledgements}
\begin{acknowledgments}
This work was supported by the United States Air Force Office of Scientific Research (AFOSR) through grant FA9550-12-1-0499, the European Research Council (Synergy Grant 609920: AXSIS), the excellence cluster ``The Hamburg Centre for Ultrafast Imaging-Structure, Dynamics and Control of Matter at the Atomic Scale'' of the Deutsche Forschungsgemeinschaft (CUI, DFG-EXC1074), and the Accelerator on Chip (ACHIP) Program funded by the Gordon \& Betty Moore Foundation. 
\end{acknowledgments}

\section*{Correspondence}
Correspondence can be directed to either P.D. Keathley (pdkeat2@mit) or F.X. K\"{a}rtner (franz.kaertner@cfel.de).

\pagebreak
\begin{widetext}
\begin{center}
\textbf{\large Supplemental Material: Examining the Transition from Multiphoton to Optical-Field Photoemission From Silicon Nanostructures}
\end{center}
\end{widetext}

%%%%%%%%%% Prefix a "S" to all equations, figures, tables and reset the counter %%%%%%%%%%
\setcounter{equation}{0}
\setcounter{figure}{0}
\setcounter{table}{0}
\setcounter{page}{1}
\makeatletter
\renewcommand{\theequation}{S\arabic{equation}}
\renewcommand{\thefigure}{S\arabic{figure}}
\renewcommand{\bibnumfmt}[1]{[S#1]}
\renewcommand{\citenumfont}[1]{S#1}
%%%%%%%%%% Prefix a "S" to all equations, figures, tables and reset the counter %%%%%%%%%%

\section*{Experimental Notes}

The following subsections highlight experimental details not included in the main text.    

\subsection*{Time of Flight Spectrometer}

The TOF spectrometer consisted of a retarding voltage screen, field free drift tube, and multichannel plate detector. The resolution at 10 eV was measured to be $\approx 40$ meV. The TOF exterior was grounded along with the surrounding chamber, and the cathode to anode spacing was roughly 3 mm.  All measurements were performed in a vacuum of approximately $1\times10^{-8}$ Torr.

\subsection*{Tip Conditioning}

Before performing experiments, we conditioned the tips with higher, sub damage threshold laser intensities and current densities (similar to the method described in~\cite{S_keathley_strong-field_2012}). The tips were exposed to an unenhanced peak intensity of $\approx1\times10^{12}$ W/cm${}^2$ for roughly 15 minutes until the current yield stabilized.  Comparable high field, high current density conditioning methods are a well accepted means of removing the surface oxide and contaminants \textit{in situ}~\cite{S_schwoebel_high_2005,S_hajra_field_2003,S_busta_current_1994}.

\section*{Calculation Notes}

The following subsections highlight notes on calculation details not included in the main text.
\\

\subsection*{Wavefunction Normalization}

\begin{figure}[h!]
\includegraphics[width=0.45\textwidth]{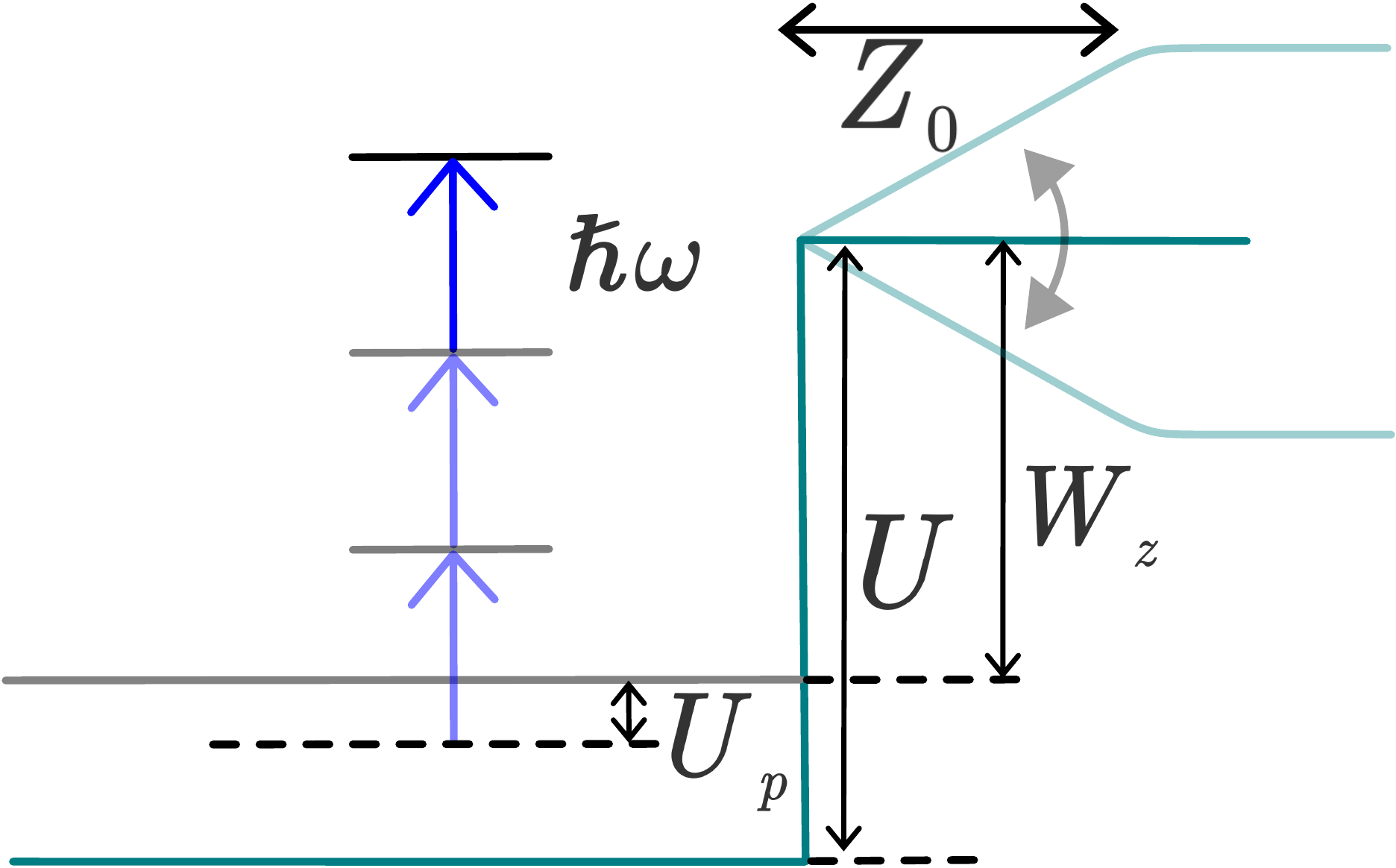}
\caption{Sketch of potential used for TDSE calculations with key parameters labeled.\label{fig:potential_sketch}}
\end{figure}

Consider the step potential of height $U$ shown in Fig.~\ref{fig:potential_sketch} before the application of an electric field to the surface.  Given an electronic plane wave with energy $W_z$ below the step propagating from the left toward the step boundary, we can write the initial state wavefunction as shown in equation~(\ref{eqn:initial_state_wavefunction}) such that $|\psi_0(z, t)| = 1$ at $z = 0$.
\begin{widetext}
\begin{equation}
\label{eqn:initial_state_wavefunction}
\psi_0(z,t) = \exp(i W_z t)\begin{cases} A\exp(ik_zz) + B\exp(-ik_zz), & z < 0 \\
\exp(-\sqrt{2 W_z} z), & z \geq 0
\end{cases}
\end{equation}
\end{widetext}

Solving for $A$ we find that $A = (i k_z - \sqrt{2W_z})/2 i k_z$.  We then have that the probability current incident on the step boundary is $(k_z^2 + 2 W_z)/4 k_z$.  Taking $\psi_0$ as the initial state for both the TDSE and TDPT, we find $C(W_z, \mathcal{E}_0)$ by integrating over the total probability density leaving the step boundary after excitation with the optical pulse.  We then find the emission probability, $\Gamma(W_z, \mathcal{E}_0)$, to be
\begin{equation}
\label{eqn:gamma}
\Gamma(W_z, \mathcal{E}_0) = \frac{4 k_z}{k_z^2 + 2 W_z}C(W_z, \mathcal{E}_0) \mbox{.}
\end{equation}

Converting everything back into energy dependent terms, we find that $\Gamma(W_z, \mathcal{E}_0) = g(W_z)C(W_z, \mathcal{E}_0)$, where 
\begin{equation}
\label{eqn:g_funct}
g(W_z) = \frac{2\sqrt{2(U - W_z)}}{U}\mbox{.}
\end{equation}

%%%%%%%%%%%%%%%%%%%%%%%%%%%%%%%%%%%%%%%%%%%%%%%
% Potential Sketch
%
% Sketch of the potential used for the TDSE
% calculations.
%%%%%%%%%%%%%%%%%%%%%%%%%%%%%%%%%%%%%%%%%%%%%%%

\subsection*{TDSE Calculations}

We numerically solved the one-dimensional TDSE with the time-dependent potential sketched in Fig.~\ref{fig:potential_sketch}. The numerical solution was based on a Crank-Nicolson scheme with discrete transparent boundary conditions (DTBCs); these specialized boundary conditions make the edges of the computational domain behave as transparent windows, i.e. the wavefunction does not reflect from the boundaries. Specifically, we implemented inhomogenous DTBCs that support exterior time-varying potentials (for further details on the implementation of such boundary conditions, see Ref.~\cite{S_antoine_review_2008}).

\subsection*{TDPT Calculations}

Starting with an initial state at $W_z$ as described in the text, and utilizing the strong-field approximation, we can approximate the amplitude of an outgoing wavepacket as
\begin{align}
\label{eqn:one_dimensional_amplitude}
M_{p_z}=&{ }-i \int\mathrm{d}\tau 
\exp \bigg \lbrace{i \bigg (S_{p_z}(\tau) + W_z \tau \bigg)} \bigg\rbrace \mathcal{E}_0\mathcal{E}_n(\tau) \times \nonumber \\
& \frac{1}{\sqrt{2\pi}}\int_{0}^{\infty} \mathrm{d}z e^{-i(p_z + A(\tau))z}e^{-\alpha(W_z) z}z \mbox{,}
\end{align}
where $A(\tau)$ is the vector potential of the optical field, $p_z$ the outgoing momentum in $z$, $\alpha(W_z) = \sqrt{2W_z}$ is the decay rate of the state in the vacuum, $\mathcal{E}_0$ the peak electric field, $\mathcal{E}_n(\tau)$ the normalized electric field waveform, and $S_{p_z}(t) = \int_{}^{t} \mathrm{d}\tau (p_z + A(\tau))^2/2$ is the action.  It is important to note that in equation~(\ref{eqn:one_dimensional_amplitude}) we have ignored rescattering (i.e. we are only dealing with direct electrons).  

The justification for only performing this integral over the vacuum half-space (for $z > 0$) is due to the significant reduction of field strength inside the emitter (for Si there is an $\approx 13.6\times$ reduction in electric field at the semiconductor/vacuum interface). 

Integrating over the probability density for $p_z > 0$, we find that   
\begin{equation}
\label{eqn:gamma}
C(W_z, \mathcal{E}_0) = \int_{0}^{\infty} \mathrm{d}p_z |M_{p_z}(W_z, \mathcal{E}_0)|^2 \mbox{.}
\end{equation}

\subsection*{Effective Mass}

While the inclusion of effective mass in the TDPT calculations had an effect on the outgoing emittance, it was found that it had a negligible impact on the total emission curve after spatial integration.  This, combined with recent work bringing into question the concept of effective mass at such short time scales~\cite{S_chang_observing_2014}, justified its omission from the calculations presented in the text.  

%%%%%%%%%%%%%%%%%%%%%%%%%%%%%%%%%%%%%%%%%%%%%%%
% Fixed Intensity Ground State Contributions
%
% Shows how resonance blurring occurs due to 
% a broad ground state energy distribution.
%%%%%%%%%%%%%%%%%%%%%%%%%%%%%%%%%%%%%%%%%%%%%%%
\begin{figure}[h!]
\includegraphics{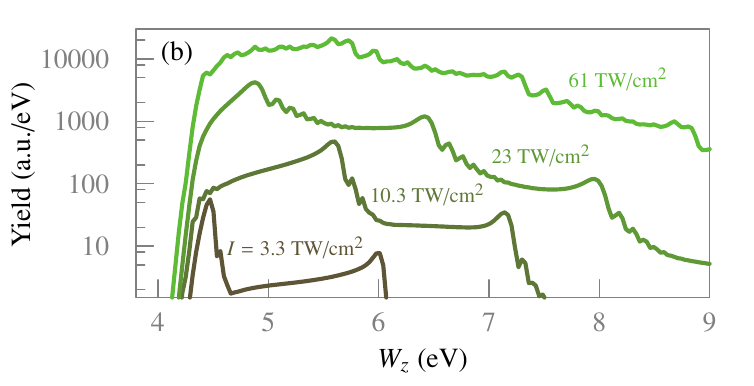}
\caption{Plot of electron yield contributions for select peak intensity values as a function of $W_z$. \label{fig:gs_fixed_I}}
\end{figure}

\section*{Ground State Contributions at Fixed Intensity Values}

In Fig.~\ref{fig:gs_fixed_I}, electron emission contributions ($F(W_z)\Gamma(W_z, \mathcal{E}_0)$) for the ideal metal at select intensities are plotted as a function of $W_z$.  For lower intensities in the MP regime, peaks form near threshold crossings, which are separated by $\hbar\omega$ ($1.55\text{\ eV}$). Furthermore, the peaks shift with increasing intensity.  To see why this is the case, we can examine the transition matrix element, finding that
\begin{align}
\label{eqn:transition_matrix_element}
\bigg|\int_{0}^{\infty}& \mathrm{d}z e^{-i (p_z + A(\tau)) z}e^{-\alpha(W_z) z}z\bigg|^2 \nonumber \\
& = \frac{1}{{\bigg [ (p_z + A(\tau))^2 + \alpha(W_z)^2 \bigg ] }^2} \mbox{,}
\end{align}
which monotonically increases with decreasing $p_z$.  Noting that for MP emission, the dominant momenta are at $p_z(n) = \sqrt{2n\hbar\omega - 2W_z - 2U_p}$ such that $n\hbar\omega > W_z + U_p$, we see that the emission will peak every time a threshold crossing occurs, i.e. when $n\hbar\omega = W_z + U_p$, and that these peaks will shift as intensity, and thus $U_p$, increases.  

For higher intensities in the OF emission regime, these peaks start to broaden into a continuum as a function of $W_z$, and the emission reaches much more deeply into the ground state distribution.  This explains the increased emission rate in the OF emission regime for the ideal metal when compared to Si.

%Supplemental Bibliography
%\begin{thebibliography}{1}
%  \input{manuscript.bbl}
%\end{thebibliography}
%\bibliography{bibliography}
\bibliographystyle{plain}

\end{document}